\documentclass{aa} 
\usepackage{graphicx}
\usepackage{txfonts}
\newcommand{\integral}{{\textit{INTEGRAL}}}
\newcommand{\xte}{{\textit{RXTE}}}
\newcommand{\sax}{{\textit{Beppo\-SAX}}}
\newcommand{\swift}{{\textit{SWIFT}}}
\newcommand{\gro}{{\textit{CGRO}}}

\newcommand{\ee}{e$^\pm$}
\newcommand{\msun}{{{\rm M}_{\sun}}}

\begin{document}

  \title{An intense state of hard X-ray emission of Cyg~X-1 observed by {\textit{INTEGRAL}} coincident with TeV measurements}
\titlerunning{An intense state of hard X-ray emission of Cyg X-1}

  \author{J. Malzac
      \inst{1}
     \and
     P. Lubi\'nski 
     \inst{2,3}
     \and
     A. A. Zdziarski
     \inst{2}
     \and 
     M. Cadolle Bel
    \inst{4}
     \and 
     M. T\"urler
     \inst{3,5}
     \and
     P. Laurent
     \inst{6}
     }
  \offprints{J. Malzac}
  \institute{Centre d'Etude Spatiale des Rayonnements (CESR), OMP, UPS, CNRS; 9 Avenue du Colonel Roche, BP44346, 31028 Toulouse Cedex 4, France \\
       \email{malzac@cesr.fr}
      \and
       Centrum Astronomiczne im.\ M. Kopernika, Bartycka 18, 00-716 Warszawa, Poland  
       \and
       ISDC Data Centre for Astrophysics, Chemin d'Ecogia 16, 1290 Versoix, Switzerland 
       \and 
        European Space Astronomy Centre (ESAC), Apartado/P.O. Box 78, Villanueva della Ca\~{n}ada, E-28691, Spain
\and
     Geneva Observatory, University of Geneva, ch.\ des Maillettes 51, CH-1290 Sauverny, Switzerland 
      \and
    CEA/DSM/Dapnia, CEA-Saclay, 91191 Gif sur Yvette Cedex, France
       }

  \date{Received 23 May 2008}

 \abstract
 %
  {}
  {We present \integral\/ light curves and spectra of the black-hole binary Cyg X-1 during a bright event that occurred in 2006 September, and which was simultaneous with a detection at 0.15--1 TeV energies by the MAGIC telescope.}
  {We analyse the hard X-ray emission from 18 to 700 keV with the \integral\/ data taken on 2006 September 24--26 by the IBIS and SPI instruments. These data are supplemented with \xte\ All Sky Monitor data at lower energy. We present the light curves and fit the high energy spectrum with various spectral models.}
  {Despite variations in the flux by a factor of $\sim$2 in the the 20--700 keV energy band, the shape of the energy spectrum remained remarkably stable. It is very well represented by an e-folded power law with the photon index of $\Gamma\simeq 1.4$ and a high energy cut-off at $E_{\rm c}\simeq 130$--140 keV. The spectrum is also well described by thermal Comptonisation including a moderate reflection component, with the solid angle of the reflector of $\sim 0.4\times 2\pi$. The temperature of the hot Comptonising electrons is $kT_{\rm e}\sim 70$ keV and their Thomson optical depth is $\tau\sim 2.5$. These spectral properties are typical of those observed in the low/hard state. This shows that Cyg X-1 may stay in the low hard state at least up to the flux level of 2 Crab, which corresponds to $\sim$2--3 \% of the Eddington luminosity. It is the first time a persistent high-mass black-hole binary is observed at a few percent of the Eddington luminosity with a stable low/hard state spectrum over a period of a few days. Such a bright hard state has so far been observed only during the rising phase of transient low-mass black-hole binaries.  The TeV detection coincides with the peak of a small X-ray flare just after a very fast rise in hard X-ray flux. In contrast, the source remained undetected by MAGIC at the peak of a larger X-ray flare occurring one day later and corresponding to the maximum of the X-ray luminosity of the whole outburst. We do not find any obvious correlation between the X-ray and TeV emission.}
 {}

  \maketitle

 \begin{figure}[!h]
  \centering
    \includegraphics[width=\columnwidth]{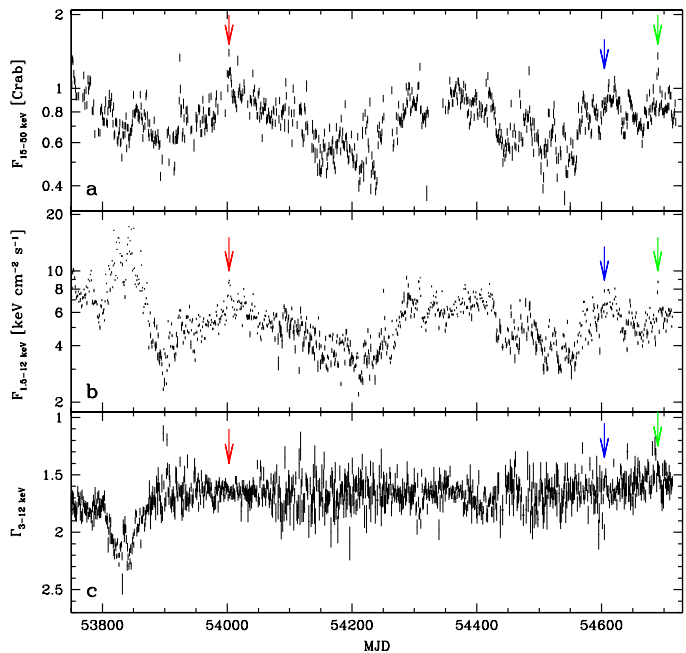}
   \caption{ Long term hard X-ray light curves of Cyg X-1. The {\it SWIFT}/BAT (15--50 keV) and the \xte/ASM (1.5--12 keV) light curves are displayed in panel (a) and (b) respectively. The BAT Crab unit is 0.221 cm$^{-2}$ s$^{-1}$, and the conversion of the ASM counts into physical flux was done following Z02. Panel (c) shows the evolution of the photon spectral index based on the 3--5 and 5--12 keV bands, as described in Z02. The red arrows at MJD 54003 point at the time of the \integral\/ observation described in this paper. The blue arrows at MJD 54604 indicate the event reported by Neronov et al.\ (2008), and the green arrows show the recent short flare peaking at MJD 56900.
}
     \label{fig:batasm}
  \end{figure}

\section{Introduction}
\label{intro}

Cyg X-1 is the prototype of black hole binaries. Since its discovery in 1964 (Bowyer et al.\ 1965), it has been intensively observed by all the high energy instruments, from soft X-rays to $\gamma$-rays. It is a persistent source, powered by accretion onto a black hole from a massive companion, HDE~226868, most likely via a focused wind. The value of the mass of the black hole is subject to controversy, it is in the range of $M_{\rm X}\simeq (5$--$15) \msun$ according to Herrero et al.\ (1995) or $M_{\rm X}\simeq (14$--$27) \msun$ (Zi{\'o}{\l}kowski 2005; see also Gies \& Bolton 1986). The distance, $D$, is most likely within $D\simeq 2.1\pm 0.2$ kpc (Zi{\'o}{\l}kowski 2005 and references therein); hereafter we adopt $D=2$ kpc.

This source is most often observed in the so-called low/hard state, characterised by a relatively low flux in the soft X-rays ($\sim$1 keV) and a high flux in the hard X-rays ($\sim$100 keV). In the hard state, the high energy spectrum can be roughly described by a power-law with photon spectral index, $\Gamma$, varying in the range $\sim$1.4--2, and a cut-off at a
characteristic energy, $E_{\rm c}$, above a hundred keV or so (e.g., Gierli{\'n}ski et al.\ 1997). Occasionally, the source switches to the high/soft state. The high-energy power law is then much softer ($\Gamma \ga 2.4$) and the bolometric luminosity is dominated by a thermal component peaking at a few keV (e.g., Gierli\'nski et al.\ 1999). In recent years, Cyg X-1 also often appeared in intermediate states in which the source exhibits spectra that are intermediate between the two canonical spectral states (relatively soft hard X-ray spectrum with $\Gamma \sim 2.0$--2.3) and with a strong flaring activity, see, e.g., Malzac et al.\ (2006). 

\integral\/ is regularly observing Cyg X-1 during monitoring observations of the Galactic Plane. On 2006 September 24--26, the source was at its highest hard X-ray level since the launch of \integral\/ in 2002 October (T\"{u}rler et al.\ 2006). During the two-day observation, the flux of the source varied significantly and reached values $>2$ Crab. Further \integral\/ observations of Cyg X-1 taken on 2006 September 30 indicated that the source returned to a more usual state at a level of $\sim$1 Crab in the 20--40 keV band and 1.3 Crab in the 40--80 keV band. 

The peak fluxes measured during 2006 September 24--26 were among the highest daily fluxes observed by the \gro\/ Burst and Transient Source Experiment (BATSE) from 1991 to 2000 in the 100--300 keV band and exceeded all of them in the 20--100 keV band (Zdziarski et al.\ 2002, hereafter Z02). Similar high fluxes were also visible in the {\it SWIFT\/} Burst Alert Telescope (BAT) and \xte\/ All Sky Monitor (ASM) light curves. { Fig.~\ref{fig:batasm} shows the long term evolution of the daily fluxes observed by the BAT\footnote{http://swift.gsfc.nasa.gov/docs/swift/results/transients/CygX-1} and ASM. The event detected by \integral\/ coincides with the highest flux ever measured by BAT (15--50 keV) since the launch of \swift\/ in 2004. The evolution of the X-ray spectral index, shown in Fig.~\ref{fig:batasm}(c), demonstrates that Cyg X-1 was then in the hard spectral state, which started around MJD 53900, and has lasted since then till now (MJD 54720). The ASM (1.5--12 keV) fluxes were also among the brightest ones in the hard state. We refer to Albert et al.\ (2007, hereafter A07) for additional close-ups of the BAT and ASM light curves around the time of the \integral\/ event.} Based on data from the ASM, the X-ray flare\footnote{A flare is a change in intensity of the source with, typically, a rising phase followed by a decay phase. It is convenient and customary to describe the X-ray variability of X-ray binaries in terms of a superposition of flares of various amplitudes and time scales. We stress that this description does not presuppose any physical mechanism for the origin of this variability, nor that each of these flares is genuinely related to an individual localised physical event.} lasted about three days, about MJD 54002.0--54005.0  (see figure 4 of A07).

{ Even brighter flares have already been reported (Stern et al.\ 2000; Golenetskii et al.\ 2003; Gierli\'nski \& Zdziarski 2003), however these events were all of much shorter duration (ms to ks rather than days). The 2006 event represents the first observation of a persistent black hole binary in a stable hard state at such levels of intensity over such a long time.} This event is also particularly interesting with regards to the simultaneous, and the first ever, detection of Cyg X-1 at TeV energies by the MAGIC telescope on 2006 September 24 (MJD 54002.928--982, A07). The 0.1--1 TeV luminosity during the MAGIC observation was $\simeq 2.2 \times 10^{34}$ erg s$^{-1}$, whereas the X-ray flare luminosity was about $5\times 10^{37}$ erg s$^{-1}$. The TeV detection was during the orbital phases of 0.90--0.91 (using the ephemeris of LaSala et al.\ 1998; Brocksopp et al.\ 1999), with the black hole behind the companion.  Also, as pointed out by Poutanen et al.\ (2008) and Rico (2008), Cyg X-1 was then close to the the peak of the long-term, superorbital, flux modulation (seen both in radio and X-rays, e.g., Lachowicz et al.\ 2006).

This paper presents a detailed analysis of the \integral\/ light curves and spectra obtained during this bright event. 

 \begin{figure*}[!th]
  \centering
  \includegraphics[width=\textwidth]{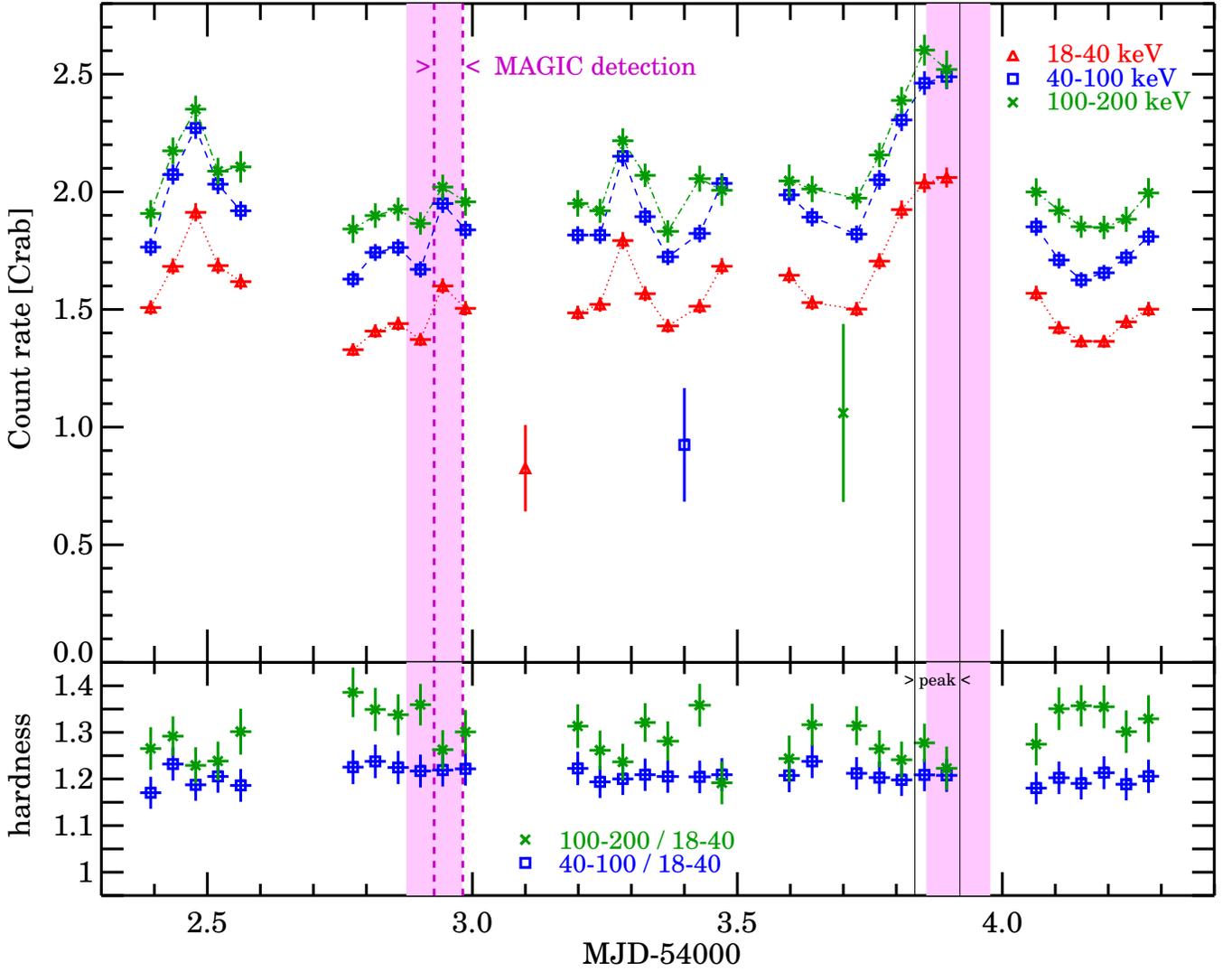}
   \caption{IBIS/ISGRI light curves during rev.\ 482 (MJD 54002.4--54004.3) The upper panel shows the light curves in the 18--40 keV (triangles), 40--100 keV (squares) and 100--200 keV (stars) bands. Each point represents the time averaged flux over a science window (a pointing of duration of $\sim$2 ks). For clarity, the data points of the 18--40 keV, 40--100 keV and 100-200 keV light curves are connected by dotted, dashed and dash-dotted lines, respectively. The three unconnected symbols (triangle, square and star), around 1 Crab, show the historical average IBIS/ISGRI fluxes of Cyg X-1 in the respective energy bands. The average fluxes were calculated using all the pointings with Cyg X-1 in the field of view since the launch of \integral. The long vertical lines associated with these symbols show the amplitude of variability (rms) of the historical \integral\/ light curves. The bottom panel shows the hardness ratios of the count rate in the 40--100 keV band to that in the 18--40 keV band (squares), and the corresponding 100--200 keV to 20--40 keV ratio (stars). The dark vertical stripes indicate the times while the source was observed with the MAGIC telescope. The thick vertical dashed lines mark the start and the end of the MAGIC pointing in which Cyg~X-1 was detected at TeV energies, the other pointings showed no detection. The two vertical solid lines enclose the two science windows corresponding to the peak spectrum shown in Fig.\ \ref{fig:spi_cutoffpl}. }
     \label{fig:splc}
  \end{figure*}

\section{Data analysis}
\label{data}

We have analysed the data corresponding to the rev.\ 482 (MJD 54002.4--54004.3). During this revolution, \integral\/ was performing a Galactic Plane scan observation, and for this reason Cyg~X-1 was always seen at least $6\degr$ from the centre of field of view. Therefore, the data are not optimal. The effective exposure was decreased by a factor of $\sim$3 when compared to the standard dithering observation of the source. As Cyg~X-1 was always outside the JEM-X field of view and no other X-ray telescope pointed at the source during this event, we lack spectral information at $\la 18$ keV, except for the ASM fluxes in three bands within 1.5--12 keV, see Section \ref{results}. In total, there were 51 pointings in rev.\ 482 when Cyg~X-1 was observed at $<18\degr$ off-axis. However, to avoid systematic errors appearing for large off-axis observations we finally chose only 31 pointings where the source illuminated at least 10\% of the ISGRI detector surface (with the off-axis angle between $6\degr$ and $13\degr$).

The IBIS/ISGRI and SPI data were reduced with the standard software of the Off-Line Scientific Analysis package (OSA, v.\ 7.0, 2007 September) released by the \integral\/ Science Data Centre (ISDC, Courvoisier et al.\ 2003). For ISGRI data analysis, we used the default OSA 7.0 settings and default response files. In case of SPI the default flat field option was chosen for the background modelling, with the constant background interval set to 5 pointings. For both instruments, the input source catalog contained all strong objects in the Cyg X-1 field, namely Cyg X-1, KS 1947+300, 4U 1957+115, QSO B1957+405, EXO 2030+375, Cyg X-3, SAX J2103.5+4545. The selection was based on the ISGRI mosaic image. 

IBIS/PICsIT data analysis was performed with the non-standard tools developed
for count rate extraction from spectral-imaging, SINGLE events data
(Lubi\'nski 2008). Basic elements of that method are: direct
handling of Poisson probability density functions for both background map
normalization and source count rate, careful modelling of the background
variability and energy-dependent pixel illumination functions. This new
approach allowed for a detection of several weaker sources and improved
the results for stronger sources when compared to the standard OSA PICsIT
software results (Lubi\'nski 2008). Usually PICsIT background maps are made for entire revolution but when some background evolution is observed, as in the case of rev.\ 482, a better result is obtained when there are more background maps. Two maps, prepared for periods before and after the end of science window number 31, allowed us to separate the background variability from the source variability. 

\begin{figure}[!th]
  \centering
    \includegraphics[width=\columnwidth]{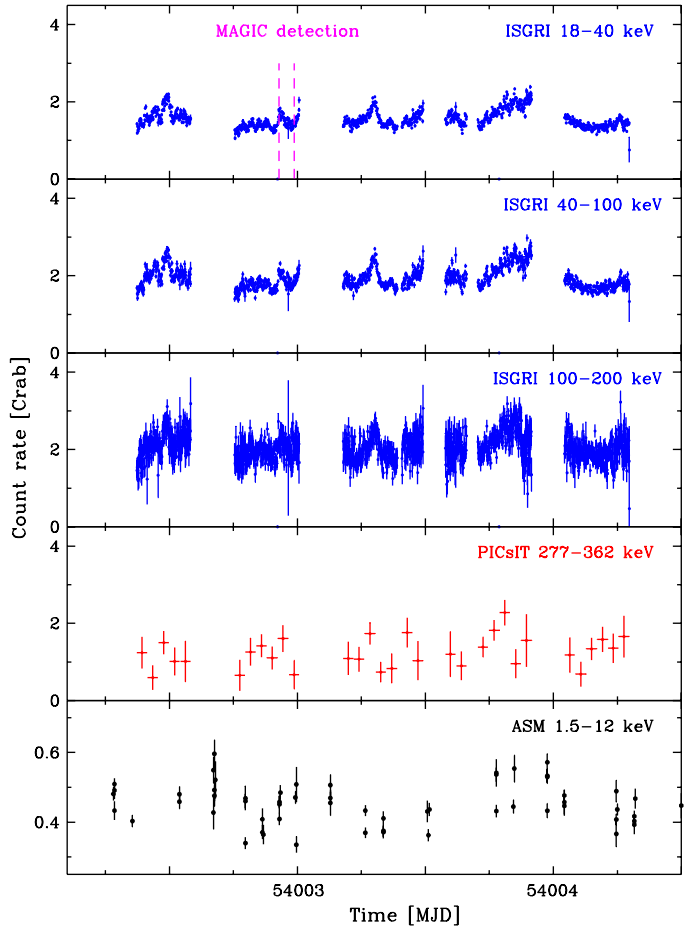}
   \caption{Light curves of Cyg X-1 during rev.\ 482. The upper panels show the IBIS/ISGRI light curves in the 18--40, 40--100 and 100--200 keV bands with a resolution of 100 s. The two bottom panels show the IBIS/PICsIT light curve in the 277--362 keV band, and the \xte/ASM 1.5--12 keV light curve. }
     \label{fig:ibisshort}
  \end{figure}

\begin{figure}[!th]
  \centering
    \includegraphics[width=\columnwidth]{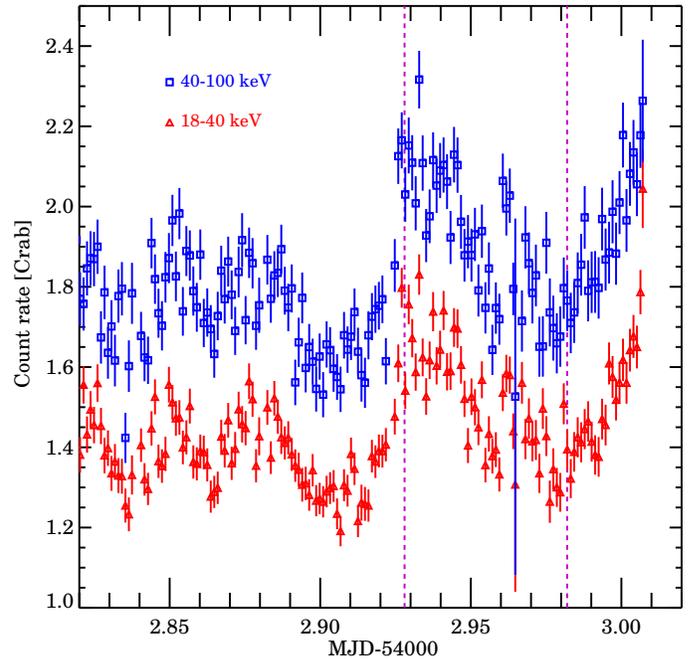}
   \caption{100-s resolution IBIS/ISGRI light curves of Cyg X-1 around the time of the MAGIC detection. The triangles and squares stand for the light curves in the 18--40 keV and 40--100 keV band respectively. The time span of the MAGIC detection is delimited by the two vertical dashed lines.}
     \label{fig:ibisshort2}
  \end{figure}

Hereafter, the light curves are presented in Crab units (see Figs.~\ref{fig:splc}--\ref{fig:ibisshort2}). For IBIS/ISGRI, we used the standard count rates provided by OSA and then define 1 Crab as 179.4, 98.3 and 17.7 s$^{-1}$ in the 18--40, 40--100 and 100--200~keV energy bands respectively. For the ASM, we used a Crab count rate of 75 s$^{-1}$. For the PICsIT, 1 Crab was defined as 3.44 s$^{-1}$ in the 277--332 keV band.   The PICsIT light curve, shown in Fig.~\ref{fig:ibisshort}, is probably the first hour-scale light curve for any source prepared for a similar energy band. Cyg X-1 is detected in practically all science windows because the corresponding $3\sigma$ upper noise limit for 3.5 ks observation, estimated on a basis of Monte Carlo simulations, is about 0.5 Crab. 
 
We have produced three spectra for each of the three instruments (IBIS/ISGRI, PICsIT and SPI). The first one is the total spectrum, i.e., averaged over the entire revolution, but with the pointings only for the offset $\leq 13\degr$ (see above). We used the same science windows (pointings) for all instruments, which resulted in 31 science windows distributed over the rev.\ 482 with total exposure $\simeq 70$ ks. The second spectrum is the average of the two pointings around the peak of the flare, (pointings 46 and 47) with the exposure of 5.2 ks, see Fig.~\ref{fig:splc}. The third one is an average of the two pointings spanning the time period MJD\ 54002.924--54003.007 coincident with the MAGIC detection at TeV energies.

  \begin{table*}
\caption{The best fit parameters of the Cyg~X-1 spectra when fitted with an e-folded power law; the e-folding energy, $E_{\rm c}$, the photon index, $\Gamma$, and the normalization at 1 keV, $K$, in units of photons keV$^{-1}$ cm$^{-2}$ s$^{-1}$. The results are shown for the spectra averaged over rev.\ 482 and for the peak of the outburst {(as indicated). In addition, the last line shows the results for the spectrum averaged over the 2 pointings that were simultaneous with the MAGIC TeV detection.}
 }
\centering
\begin{tabular}{l c c c c c c}   
\hline\hline    
 instrument   & $E_{\rm c}$  (keV)         & $\Gamma$       &      $K$            & $\chi^2/\nu$ ($\nu$) \\ 
\hline          
 {SPI}    & $139^{+10}_{-10}$ & $1.40^{+0.03}_{-0.03}$  &   1.86$^{+0.2}_{-0.2}$ &    $0.86 (22)$\\
 {SPI} (peak)   & $161^{+52}_{-34}$  & $1.51^{+0.12}_{-0.12}$  &  3.72$^{+1.56}_{-1.14}$ &  0.82 (22)\\
 {ISGRI}     & $128^{+6}_{-5}$  & $1.36^{+0.02}_{-0.02}$  & 1.42$^{+0.10}_{-0.09}$ &   1.08(48) \\
 {ISGRI} (peak)  & $113^{+10}_{-9}$ & $1.31^{+0.04}_{-0.04}$ & $1.63^{+0.21}_{-0.18}$ & 1.13 (48)  \\
 {SPI}+{ISGRI}+{PICsIT} & $131^{+5}_{-5}$ & $1.37^{+0.02}_{-0.02}$  &     1.66$^{+0.10}_{-0.10}$  & 0.97 (78)\\ 
 {SPI}+{ISGRI}+{PICsIT} (peak)& 117$^{+10}_{-8.5}$ & 1.33$^{+0.04}_{-0.04}$  &    2.08$^{0.24}_{-0.22}$ & 1.16 (75) \\
 {SPI}+{ISGRI}+{PICsIT} (TeV detection) & 111$^{+12}_{-10}$ & 1.34$^{+0.05}_{-0.05}$ & 1.25$^{+0.21}_{-0.18}$ & 0.80 (45) \\
\hline         
\end{tabular}
\label{tab:efold}
 \end{table*}

The spectral analysis was performed under XSPEC v.\ 11.3 (Arnaud et al. 1996).
The energy range used was 18--300 keV for the IBIS , 24--700 keV for the SPI, and 277--632 keV for the PICsIT. A 1.5\% systematic error was added in quadrature to the statistical errors of the IBIS/ISGRI spectra. Moreover, in order to account for uncertainties in the cross-calibration when simultaneously fitting spectra from different instruments, a multiplicative constant was added in the spectral fits to each instrument data set: it was set free for IBIS and frozen to 1 for SPI. The resulting best-fit value for these constant was around $\simeq 0.9$ for ISGRI. For PICsIT the normalisation constant is not very well constrained by the data. We found values in the range 0.8--1.2 depending on the data set and model used. 

 \begin{figure}
  \centering
  \includegraphics[width=\columnwidth]{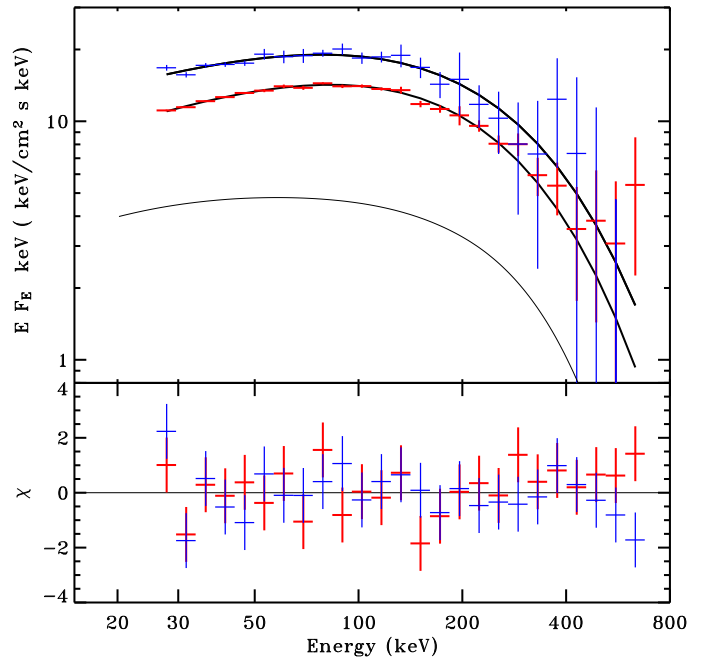}
   \caption{SPI spectra at the outburst peak (blue) and averaged over the rev.\ 482 (red), fitted by the e-folding power-law model (see Table \ref{tab:efold}). For comparison, we also show the best-fit model of a typical hard-state spectrum (period 1 of Cadolle Bel et al.\ 2006). }
     \label{fig:spi_cutoffpl}
  \end{figure}

  \begin{figure}
  \centering
  \includegraphics[width=\columnwidth]{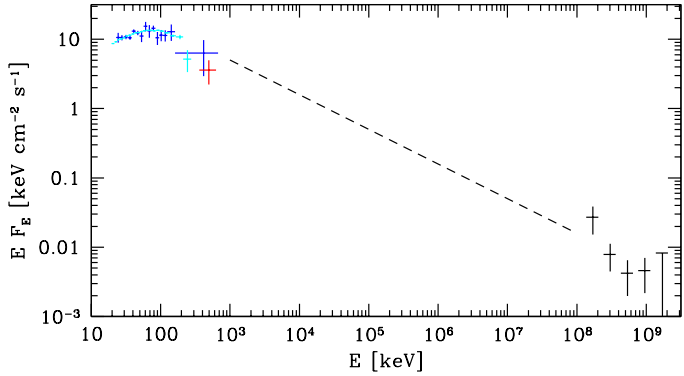}
   \caption{Simultaneous \integral/MAGIC spectrum of Cyg~X-1. The SPI (blue), ISGRI (cyan) and PICsIT (red) spectra are averaged over the time of the TeV detection. For clarity, all the \integral\/ data were rebinned. The ISGRI data were normalised to the SPI spectrum (the ISGRI flux was divided by 0.96) as determined from the fit with a cut-off powerlaw model (see Table~\ref{tab:efold}). The black crosses show the simultaneous MAGIC spectrum (A07). The dashed line is a power-law with photon index $\Gamma=2.5$ connecting the two data sets.}
     \label{fig:spintegralmagic}
  \end{figure}

 \begin{figure}
  \centering
  \includegraphics[width=\columnwidth]{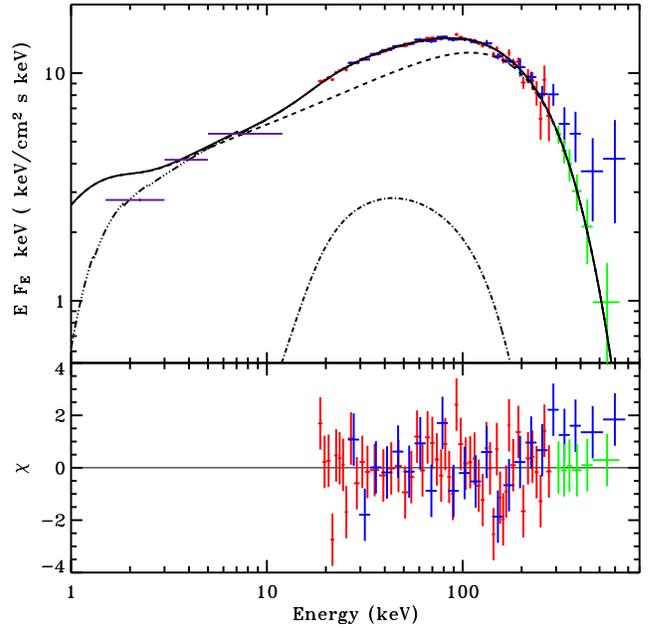}
 \caption{SPI (blue), ISGRI (red) and PICsIT (green) spectra averaged over rev.\ 482 fitted with the model consisting of thermal Comptonisation (dashed curve) and Compton reflection (dot-dashed curve) model, see Table \ref{tab:compps}. The solid curve gives the sum. { The triple-dot-dashed curve shows the same model but taking into account the standard absorption column density toward the source ($N_{\rm H}=6\times 10^{21}$ cm$^{-2}$).} For clarity, all the data were rebinned and normalised to the SPI spectrum (the ISGRI and PICsIT fluxes were divided by 0.889 and 1.18 respectively). The three dark horizontal lines below 12 keV show the average ASM energy fluxes over MJD 54002.4--54004.3 (those data were not taken into account in the fit procedure).}
 \label{fig:allint}
 \end{figure}

\section{Results}
\label{results}

The IBIS/ISGRI light curves of rev.\ 482 are shown in Fig.~\ref{fig:splc}. { As shown on this figure, during the whole observation the measured fluxes were a factor of 2 above the historical average IBIS/ISGRI fluxes and also well outside the range of the average variability amplitude. The light curves} can be decomposed into an underlying approximately constant component at $\simeq$1.5 Crab (18--40 keV), $\simeq$1.8 Crab (40--100 keV) and $\simeq$1.9 Crab (100--200 keV), and, on top of it, three main flares lasting for several hours and reaching fluxes $>2$ Crab. The strongest outburst and reached the maximum of about 2 Crab (18--40 keV), 2.4 Crab (40--100 keV) and 2.5 Crab (100--200 keV) in a pointing on 2006 September 25 20:58--21:57 UT. This main flare lasted for 10 hours (estimated from the time interval between the 2 minima in the light curve before and after the peak).

The times when MAGIC was pointing at the source and the time of the TeV detection are also marked on this figure. Strikingly, MAGIC did not detect Cyg~X-1 at the peak of the 10-hour outburst (A07), but one day earlier, when the the hard X-ray flux was close to the average level of our observation. 

The 40--100 keV to 18--40 keV (hereafter HR1) and the 100--200 keV to 18--40 keV (hereafter HR2) hardness ratios are shown in the bottom panel of Fig.\ \ref{fig:splc}. Both hardness ratios are remarkably stable: the spectrum remains impressively constant in spite of the relatively strong flux fluctuations. Nevertheless, although we do not detect any fluctuation of HR1, we find small ($<$15\%) but significant variations of HR2. As seen on Fig.\ \ref{fig:splc}, the HR2 tends to be anti-correlated with the flux. This is confirmed by a Spearman rank test showing that HR2 is anti-correlated with the flux in the 40--100 keV energy band with probability of $2\times 10^{-6}$ that this correlation occurred by chance. This behaviour is similar to that shown during an \integral\/ observation during the normal hard state, see fig.\ 3 in Bazzano et al.\ (2003), and, on longer time scales, by the BATSE, see fig.\ 6a in Z02. The origin of this effect was modelled by Z02 as due variability of the total luminosity in a self-consistent thermal Comptonization model. It is also worth noting that HR2 was at a very similar level during the MAGIC detection and at the peak of the outburst. Thus the TeV emission seems unrelated to the hardness ratio. 

Fig.\ \ref{fig:ibisshort} shows the IBIS/ISGRI light curves with a higher time resolution, 100 s, which reveal a strong flaring hard X-ray activity also occuring on shorter time scales. Using the full light curve to compute the rms amplitude of the variability, we find that the Poisson noise corrected rms variability amplitude is of 23, 27 and 23\% in the 18--40, 40--100 and 100--200 keV bands respectively.  Statistical noise contributes at a level of 3.2, 5.5 and 51\%, respectively, to the variance of the plotted light curves. 

The intrinsic variability appears, however, dominated by the longer time scales. 
In order to evaluate the amplitude of variability occurring on time scales shorter than one hour, we computed the rms on segments of duration 3400 s, which removes the longer time scales. The average, noise corrected, rms amplitude is then reduced to 7\% in all three energy bands. The noise contributes at a level of 10, 16 and 73\% to the observed variance (i.e. most of the short time scale variability of the 100--200 keV band seen on Fig.\ \ref{fig:ibisshort}, is actually due to statistical noise). 

A close-up of the light curves around the time of the MAGIC detection is presented on Fig.\ \ref{fig:ibisshort2}. The start of the TeV emission coincided {with the peak of an X-ray flare}, while there was no TeV detection for 2.5 hours before. However the MAGIC detection could also be seen (from Fig.\ \ref{fig:ibisshort} and\ \ref{fig:ibisshort2}) as preceeding the sharp rise of a flare not fully covered by \integral\/ observations. If the TeV emission is indeed a precursor and not simultaneous to the hard X-ray emission, this may help to understand the absence of TeV detection during the peak of the major 10-hour flare. { Alternatively, we note that the detection occurs just after a rapid ($<$ 5 min) increase in the 40-100 keV X-ray intensity. This brightening appears to be the most abrupt change in flux during the whole \integral\/ observation (see Fig.~\ref{fig:ibisshort}) and the TeV emission might also be associated to this fast rise in the X-ray luminosity.} In fact, the lack of coverage and the uniqueness of the TeV detection prevent us from linking firmly the TeV emission to any feature of the hard X-ray light curve. 

Fig.~\ref{fig:spi_cutoffpl} shows both the spectrum averaged over rev.\ 482 and the peak SPI spectrum. In addition it shows, as reference, the best-fit model of a typical hard-state spectrum previously observed by \integral\/ (Cadolle et al.\ 2006). Notably, it has a flux lower by a factor of $\sim$2 but a very similar spectral shape. Both SPI spectra are well represented by an e-folded power-law model characterised by a power-law photon index $\Gamma \sim 1.4$ and a cut-off energy $E_{\rm c}\sim140$ keV. The best fits parameters are shown in Table \ref{tab:efold}. Both spectra have the best fits parameters consistent with being the same within the uncertainties, confirming the absence of any strong spectral evolution during the flare. We have also performed similar fits using the ISGRI spectra and combining the SPI, ISGRI and PICsIT. The best fit parameters are shown in Table \ref{tab:efold} and are similar to those obtained using SPI alone. The small differences in the best fit parameters are due to known calibration problems. The spectrum during the outburst is consistent with a stable hard state. This shows that Cyg X-1 may stay in the hard state at least up to fluxes of 2 Crab. 

Fig.~\ref{fig:spintegralmagic} shows the spectrum of the source at the time of the MAGIC detection, together with the simultaneous TeV data (A07). Although the statistics is relatively poor, the \integral\/ spectrum appears to be very similar to the average and peak spectra (see best fit parameters in Table~\ref{tab:efold}).
 
We also fit the average spectrum with the Comptonisation model {\sc compps} (Poutanen \& Svensson 1996). In this thermal Compton model, the spectrum is computed for a homogenous spherical hot corona, inside which blackbody photons are injected and then Compton upscatered by hot Maxwellian electrons. The corona is parametrized by its electron temperature, $T_{\rm e}$, and the radial Thomson optical depth, $\tau=n_{\rm e}\sigma_{\rm T} H$, where $n_{\rm e}$ is the electron density, $\sigma_{\rm T}$ is the Thomson cross section, and $H$ the radius of the sphere. The temperature of the seed photons is kept fixed at 0.3 keV. This model also accounts for the reflection of the hard X-ray radiation illuminating the disc and forming an additional reflection bump peaking at $\sim$30 keV (Magdziarz and Zdziarski 1995), for which we assume the inclination of $45\degr$. The strength of the reflection component is quantified using its relative amplitude $R$ with respect to the case of the X-ray emission of an optically-thin isotropic source reflected by an infinite slab of cold material. The abundances of the reflector are solar and the reflecting material is neutral. 
 
The best fit parameters are shown in Table \ref{tab:compps}. We fit the ISGRI and SPI spectra independently as well as the combined ISGRI/PICsIT/SPI spectrum. In the fits, we find $kT_{\rm e}\sim 70$ keV and $\tau\sim 2.5$. The presence of a reflection component is required. The fits performed without it are poor, with $\chi^2/\nu>2.3$, whereas its inclusion improves the fits dramatically, leading to $\chi^2 /\nu \simeq 1$. The combined ISGRI/SPI/PICsIT spectrum, shown in Fig.~\ref{fig:allint}, requires a moderate reflection strength of $R\sim 0.4$. Overall, the parameters of the thermal Compton and reflection model are similar to those usually obtained in the low/hard state (see, e.g., Done et al. 2007). In Fig.~\ref{fig:allint} we also show the average ASM energy fluxes over MJD 54002.4--54004.3, converted from the count rates using the response matrix of Z02. These fluxes turn out to be consistent with the low energy extrapolation of the best fit model of the \integral\/ spectrum, confirming the validity of the Comptonization model (the extrapolation of the best fit e-folded power law model does not provide such a good agreement with the ASM data). 

Non-thermal high energy components in excess of thermal Comptonisation models have been reported in the hard state of Cyg X-1 (McConnell et al.\ 2000, 2002, Cadolle Bel et al. 2006), and, e.g., the low-mass X-ray binary GX 339--4 (Wardzi\'nski et al.\ 2002; Joinet et al.\ 2007). Such a hard tail is usually attributed to the presence of a small fraction of non-thermal electrons in the hot comptonising medium. In the present data the presence of such an excess is suggested by the shape of the SPI residuals above 300 keV (see in Fig.\ \ref{fig:allint}). We therefore fit again the combined spectrum shown in Fig.\ \ref{fig:allint} allowing for the presence of a non-thermal tail in the Comptonizing electron distribution. This did not lead to any sensible improvement of the $\chi^2$. This excess does not appear in the PICsIT data, but this depends on the poorly constrained normalisation factor between SPI and PICsIT. We therefore conclude that, even though it is not excluded, we do not find robust evidence for the presence of a non-thermal excess in this observation. 

\begin{table}
\caption{The best fit parameters of the Cyg X-1 spectra for thermal Comptonization; the electron temperature, $kT_{\rm e}$, the Thomson optical depth, $\tau$, and the strength of Compton reflection, $R$, relative to that from an infinite slab. The fit results with the fixed (denoted by 'f') $R=0$ are also shown.
 }
\centering
\begin{tabular}{c c c c c c c}   
\hline\hline    
 instrument   & $kT_{\rm e}$ (keV)           & $\tau$          & $R$        & $\chi^2/\nu$ ($\nu$)\\ 
\hline          
 {SPI}     & $64^{+3}_{-3}$   & $2.36^{+0.07}_{-0.06}$  &         0f             & $2.51 (22)$\\
 
 {SPI}       & $75^{+7}_{-6}$  & $2.25^{+0.22}_{-0.09}$  &    $0.40^{+0.1}_{-0.2}$      & 1.24 (21) \\
 {ISGRI}     & $55^{+3}_{-2}$   & $2.94^{+0.15}_{-0.21}$  &            0f             & 2.36(48) \\
 {ISGRI}     & $64^{+3}_{-2}$   & $2.61^{+0.12}_{-0.07}$  &  $0.26^{+0.1}_{-0.09}$ & 1.14 (47)\\
 {SPI}+{ISGRI}+{PICsIT} & $58^{+1}_{-2}$ & $2.73^{+0.14}_{-0.05}$ &   0f             & 2.54(78)\\ 
 {SPI}+{ISGRI}+{PICsIT} & $68^{+3}_{-3}$   & $2.46^{+0.07}_{-0.07}$ &  $0.30^{+0.09}_{-0.07}$ & 1.13 (77)\\
\hline         
\label{tab:compps}
\end{tabular}
 \end{table}
 
\section{Discussion}
\label{discussion}

\begin{figure}
  \centering
  \includegraphics[width=\columnwidth]{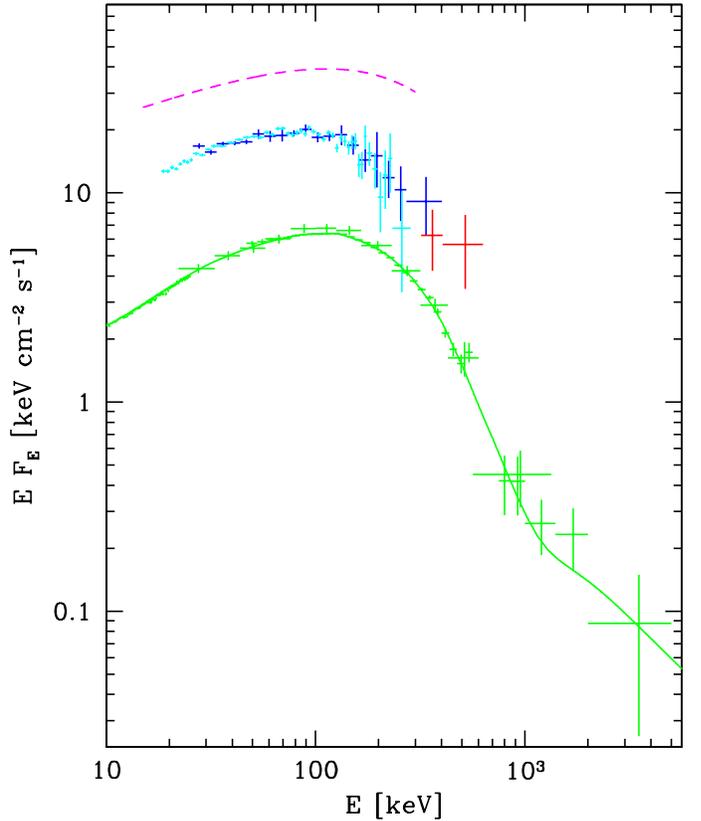}
 \caption{Comparison of the \integral\/ spectrum corresponding to the peak of the flare (the SPI data are shown in blue, ISGRI in cyan and PICsIT in red crosses; the data have been rebined for clarity), the average Cyg X-1 hard-state spectrum (McConnell et al.\ 2002) from the \gro/OSSE/BATSE/COMPTEL including a \sax\/ spectrum at low energies (the green histogram fitted by the solid curve), and the brightest previously recorded ks-time scale flare (the dashed 15--300 keV curve). For that last spectrum, we show the fit to the average over the period C of the flare 990421A (table 2 in Golenetskii et al.\ 2003). Note that the peak flux of that flare was still 1.8 times higher than that of the average spectrum shown here. For clarity the \integral\ data have been rebined and normalised to the SPI spectrum according to the best fit e-folded power-law model. The ISGRI and PICsIT fluxes were divided by 0.84 and 0.85 respectively (the PICsIT normalisation factor is poorly constrained). 
}
 \label{fig:compare}
 \end{figure}

Events similar to the luminous state presented in this paper were previously reported in Cyg X-1. Stern et al.\ (2001) discovered bright hard X-ray flares lasting $\sim$1 ks in the BATSE data. These events occured in the hard state with the peak luminosity at $>30$ keV reaching $1.6\times 10^{38}$ erg s$^{-1}$, which is an order of magnitude larger than the normal luminosity of Cyg X-1 in this state. They found two types of spectral behaviour during these outbursts, one with constant spectrum, similar to what we observed in September 2006, the other showing spectral varibility suggesting the presence of two spectral components. Golenetskii et al.\ (2003) reanalysed the archives of the BATSE, {\it Ulysses}, and the {\it Wind}/Konus experiments, and found 7 strong outbursts of duration $<$ 28 ks, both in hard and soft state, with the peak 15--300 keV luminosity similar to the one found by Stern et al.\ (2001). Fig.\ \ref{fig:compare} shows the spectrum fitted to the brighest of those flares (in the hard state) and compares it with the peak \integral/SPI spectrum, as well as with the average Cyg X-1 hard-state spectrum from McConnell et al.\ (2002). Note that the peak flux of brightest previous flare was still a factor 1.8 higher than that shown in Fig.\ \ref{fig:compare} (see fig.\ 5 and compare tables 2 and 3 in Golenetskii et al.\ 2003). The isotropic bolometric luminosity inferred from that fit and adjusted to the peak is $L\sim 3\times 10^{38}$ erg s$^{-1}$, which is $\sim 20\%(10\msun/M)$ of the Eddington luminosity, $L_{\rm E}\simeq 1.5(M/10\msun) 10^{39}$ erg s$^{-1}$ (assuming the solar abundances).

In addition, powerful sub-second flares were found by Gierli\'nski \& Zdziarski (2003; see also Zdziarski \& Gierli\'nski 2004), mostly in the hard state. The hard-state flares reached the 3--30 keV luminosity of $\sim\! 10^{38}$ erg s$^{-1}$. Their estimated bolometric luminosity was up to $\sim\! 4\times 10^{38}$ erg s$^{1}$, which is $\sim 25 \%(10\msun/M) L_{\rm E}$. We have searched for similar short time-scale flaring activity during this outburst. We examined the evolution of the SPI and ISGRI individual detector count rate on time scales of a few seconds (and down to 0.1 s with ISGRI) around the peak of the outburst but found no evidence for the presence of such short flares in our data. PICsIT spectral timing data were also checked for the presence of very short flares. With the 16 ms time resolution nothing was found in 8 energy bands between 208 keV and 2.6 MeV.

In the context of previous observations, the 2006 September event is therefore not extreme in terms of luminosity. The bolometric flux for our e-folded power law model using the SPI data is $8.5 \times 10^{-8}$ and $1.2 \times 10^{-7}$ erg cm$^{-2}$ s$^{-1}$ for the average of rev.\ 482 and its peak, respectively. This corresponds to (2.8--$3.9)\%(10\msun/M) L_{\rm E}$. These Eddington ratios are still below those seen in the soft state of Cyg X-1, $\simeq 5\%(10\msun/M) L_{\rm E}$ (Z02), unlike the case of the shorter flares, which $L$ substantially exceeded this value. 

However, the \integral\/ flare is exceptional in terms of duration. The source stayed in this bright state for several days, and the 10-hour ouburst is among the longest recorded. This suggests a relative stability of this bright state. The fact that this flare was below the flux level typical of the soft state could be related to its long duration. Were it above the soft-state level, it would, most likely, have triggered a state transtion to the soft state during the 2-day time scale, perhaps like the one described in Malzac et al.\ (2006). 

We note that another flare, indicated in Fig.~\ref{fig:batasm} by the blue arrows, was recently reported by \integral\/ (Neronov et al.\ 2008). It was, however, substantially weaker (it did not even correspond to a local flux maximum of the ASM/BAT light curves), and with a shorter time scale than the 2006 event. { On the other hand, an event with the amplitude approximately equal to the 2006 one, but still substantially shorter (of a $\sim$1-d duration), took place on MJD 54690 (2008 August 12), as shown by the green arrows on Fig.~\ref{fig:batasm}. It appears to correspond to the maximum of the superorbital cycle of Cyg X-1, currently with the period of $\sim$300 d (Rico 2008). It occured, however, 35 d later than the date predicted by Rico (2008), indicating a quasiperiodic character of the superorbital modulation.}

Several other sources like XTE~J1739--302, Vela~X-1 and 1E~1145--6141, which also exihibit relatively short time-scale (hours) flares, share with Cyg X-1 the fact of being bright X-ray binairies with a giant companion where accretion proceeds trough the wind of the companion star (Negueruela et al.\ 2006; Smith et al.\ 2006). At least some of these bright events are likely to be caused by an increased mass accretion rate triggered by the variability of the wind.  
 
The fact that the spectrum observed in this luminous state is very similar to the hard state spectrum suggests that the emission mechanism is similar. In the hard state, those spectra are usually believed to form close to the black hole in some kind of hot geometrically thick, optically thin accretion flow. Although the simplest advection dominated accretion flow solutions (Narayan \& Yi 1995) may not hold at such high luminosities, other forms of luminous hot accretion flow solutions are known to exist and to be sufficiently stable (Yuan 2001, 2004; Yuan et al.\ 2007). 

Alternatively, the emission could be produced in an outflowing, optically-thin, patchy corona atop a standard geometrically-thin, optically-thick accretion disc (Beloborodov 1999; Malzac et al.\ 2001). The possibility that the hard X-ray emission during the flare arises in a jet due to a microblazar activity as suggested by Romero et al.\ (2002) is ruled out as this model predicts a softer spectrum during bright flares. Also, the flare periodicity predicted by that model has been shown to be absent in the data (Lachowicz et al.\ 2006).
 
The \integral\/ spectrum at the time of the MAGIC detection was a typical hard state spectrum, and similar to the peak spectrum (during the non-detection). In fact, we do not find any unusual features in the X-ray light curves and spectrum around the time of the MAGIC detection, except for the high X-ray luminosity that lasted for a few days.
 
As discussed in Section\ \ref{results}, we do not find conclusive evidence for a non-thermal high energy component in the present data. However, as already mentioned, deeper exposure observations performed when the source was in the hard state have detected such a hard tail. It is usually attributed to the presence of a small fraction of non-thermal electrons in the hot comptonising medium. In the case of Cyg X-1, this non-thermal component extends at least up to a few MeV (McConnell et al.\ 2002, see Fig.\ \ref{fig:compare}). The TeV emission reported by A07 could in principle be related to such non-thermal coronal electrons. If we extrapolate the reported TeV emission down to the MeV range assuming $\Gamma=2.5$ (a characteristic non-thermal index in Cyg X-1, e.g., Gierli\'nski et al.\ 1999), we match the SPI residuals at $\sim$500 keV (Figs.\ \ref{fig:spintegralmagic}--\ref{fig:allint}). This spectral index is very close to the range of $\Gamma=3.2\pm 0.6$ measured by the MAGIC detector (A07). However, it is highly unlikely that the spectrum from $<1$ MeV up to 1 TeV is a single power law. Even if the intrinsic spectrum is a power law, reprocessing by \ee\ pair production events will change its shape. The issue of pair absorption and pair reprocessing is, however, beyond the scope of this paper and calculations of these effects will be presented elsewhere. 

Alternatively, the TeV emission could arise in shocks located in the region where the outflow originating close to the black hole interacts with the wind of the star. Such an outflow is indeed observed in Cyg X-1 in the form of the compact radio jet. Other forms of outflows such as an accretion disc wind could also be responsible for this emission. This outflow colliding with the star wind may power strong shocks accelerating the particles responsible for the TeV emission. In this scenario an increase of the wind strength would increase the mass accretion rate onto the black hole, leading to the enhanced hard X-ray emission. Simultaneously the stronger wind interacts more efficiently with the (possibly stronger) outflow, making the TeV emission detectable. Still, pair absorption in the photon field of the companion (e.g., Dubus 2006) remains to be a major issue. 

\begin{acknowledgements}
This paper is based on observations with \integral, an ESA project with instruments and science data centre funded by ESA member states (especially the PI countries: Denmark, France, Germany, Italy, Switzerland, Spain), Czech Republic and Poland, and with the participation of Russia and the USA. This research has been supported in part by the CNRS, the LEA Astrophysics Poland-France (Astro-PF) program, the Polish MNiSW grant NN203065933 (2007--2010), and the Polish Astroparticle Network 621/E-78/SN-0068/2007. JM thanks Elisabeth Jourdain for many discussions of the \integral\/ data, and the Institute of Astronomy of the University of Cambridge (UK) for hospitality during the final stage of this work. We are also grateful to Olivier Godet for advices on the {\it SWIFT}/BAT data. We thank the MAGIC collaboration for providing us with the Cyg X-1 spectrum in electronic form.

\end{acknowledgements}


\begin{thebibliography}{}

\bibitem[Albert et al.(2007)]{2007ApJ...665L..51A} Albert, J., and the MAGIC collaboration, 2007, \apjl, 665, L51 (A07)

\bibitem[Arnaud(1996)]{1996ASPC..101...17A} Arnaud, K.~A.\ 1996, 
Astronomical Data Analysis Software and Systems V, 101, 17 

\bibitem[]{bazzano03}
Bazzano, A., Bird, A. J., Capitanio, F., et al.\ 2003, \aap, 411, L389

\bibitem[Beloborodov(1999)]{1999ApJ...510L.123B} Beloborodov, A.~M.\ 1999, 
\apjl, 510, L123 

\bibitem[\protect\citeauthoryear{Brocksopp et al.}{1999}]{brock2}
Brocksopp, C., Tarasov, A. E., Lyuty, V. M., Roche, O. 1999, A\&A, 343, 861

\bibitem[Cadolle Bel et al.(2006)]{2006A&A...446..591C} Cadolle Bel, M., Goldwurm, A., Rodriguez, J., et al.\ 2006, \aap, 446, 591 

\bibitem[Courvoisier et al. 2003]{c03}
 Courvoisier, T. J.-L., Walter, R., Beckmann, V., et al.\ 2003, A\&A, 411, L53

\bibitem[Done et al.(2007)]{2007A&ARv..15....1D} Done, C., Gierli{\'n}ski, 
M., \& Kubota, A.\ 2007, \aapr, 15, 1 

\bibitem[\protect\citeauthoryear{Dubus}{2006}] {dubus06}
Dubus, G. 2006, A\&A, 451, 9

\bibitem[Gierli{\'n}ski \& Zdziarski(2003)]{2003MNRAS.343L..84G} 
Gierli{\'n}ski, M., \& Zdziarski, A.~A.\ 2003, \mnras, 343, L84 

\bibitem[Gierlinski et al.(1997)]{1997MNRAS.288..958G} Gierlinski, M., 
Zdziarski, A.~A., Done, C., et al. \ 1997, \mnras, 288, 958 

\bibitem[\protect\citeauthoryear{Gierli{\'n}ski et 
al.}{1999}]{Gie99} Gierli{\'n}ski M., Zdziarski A.~A., 
Poutanen J., Coppi P.~S., Ebisawa K., Johnson W.~N., 1999, MNRAS, 309, 496 

\bibitem[\protect\citeauthoryear{Gies \& Bolton}{1986}]{gies86} 
Gies, D.~R., \& Bolton, C.~T., 1986, ApJ, 304, 371 

\bibitem[Golenetskii et al.(2003)]{2003ApJ...596.1113G} 
Golenetskii, S., Aptekar, R., Frederiks, D., et al. 2003, \apj, 596, 1113

\bibitem[\protect\citeauthoryear{Herrero et al.}{1995}]{herrero95} 
Herrero, A., Kudritzki, R. P., Gabler, R., Vilchez, J. M., \& Gabler, A., 1995, A\&A, 297, 556

\bibitem[Joinet et al.(2007)]{2007ApJ...657..400J} Joinet, A., Jourdain, 
E., Malzac, J., et al. \ 
2007, \apj, 657, 400 

\bibitem[\protect\citeauthoryear{Lachowicz et al.}{2006}]{l06}
Lachowicz, P., Zdziarski, A. A., Schwarzenberg-Czerny, A., Pooley, G. G., 
\& Kitamoto, S. 2006, MNRAS, 368, 1025

\bibitem[\protect\citeauthoryear{LaSala et al.}{1998}]{lasala98}
LaSala, J., Charles, P.A., Sith, R. A. D., Ba{\l}uci{\'n}ska-Church, M., 
\& Church, M. J. 1998, MNRAS, 301, 285


\bibitem[Lubi\'nski (2008)]{l08}
{Lubi\'nski, P. 2008, A\&A,  submitted, arXiv:0809.0427}

\bibitem{mz99} Magdziarz, P., \& Zdziarski, A. A. 1995, MNRAS, 273, 837

\bibitem[Malzac et al.(2001)]{2001MNRAS.326..417M} Malzac, J., Beloborodov, 
A.~M., \& Poutanen, J.\ 2001, \mnras, 326, 417 

\bibitem[Malzac et al.(2006)]{2006A&A...448.1125M} Malzac, J., Petrucci , P.O., Jourdain, E., et al.\ 
2006, \aap, 448, 1125 

\bibitem[McConnell et al.(2000)]{2000ApJ...543..928M} McConnell, M.~L., Ryan, J. M.; Collmar, W., et 
al.\ 2000, \apj, 543, 928 

\bibitem[McConnell et al.(2002)]{2002ApJ...572..984M} McConnell, M.~L., Zdziarski, A. A., Bennett, K., et 
al.\ 2002, \apj, 572, 984 

\bibitem[Narayan \& Yi(1995)]{1995ApJ...452..710N} Narayan, R., \& Yi, I.\ 
1995, \apj, 452, 710 

\bibitem[Negueruela et al.(2006)]{2006ESASP.604..165N} Negueruela, I., 
Smith, D.~M., Reig, P., Chaty, S., \& Torrej{\'o}n, J.~M.\ 2006, The X-ray 
Universe 2005, 604, 165 

\bibitem[Neronov et al.(2008)]{2008ATel.1533....1N} Neronov, A., Cadolle Bel, M., Shaw, S., et al.\ 2008, ATel, 1533, 1 (errata in ATel 1536, 1)

\bibitem[Poutanen \& Svensson(1996)]{1996ApJ...470..249P} Poutanen, J., \& 
Svensson, R.\ 1996, \apj, 470, 249 

\bibitem[\protect\citeauthoryear{Poutanen, Zdziarski, 
\& Ibragimov}{2008}]{2008arXiv0802.1391P} 
Poutanen, J., Zdziarski, A.~A., \& Ibragimov, A. 2008, MNRAS, 389, 1427

\bibitem[Rico (2008)]{r08} Rico, J. 2008, \apj, 683, L55

\bibitem[Romero et al.(2002)]{2002A&A...393L..61R} Romero, G.~E., Kaufman 
Bernad{\'o}, M.~M., \& Mirabel, I.~F.\ 2002, \aap, 393, L61 

\bibitem[Smith et al.(2006)]{2006ApJ...638..974S} Smith, D.~M., Heindl, 
W.~A., Markwardt, C.~B., et al. \ 2006, \apj, 638, 974 

\bibitem[Stern et al.(2001)]{2001ApJ...555..829S} Stern, B.~E., 
Beloborodov, A.~M., \& Poutanen, J.\ 2001, \apj, 555, 829 

\bibitem[T{\"u}rler et al.(2006)]{2006ATel..911....1T} T{\"u}rler, M., Zdziarski, A. A., Laurent, P., et 
al.\ 2006, ATel, 911, 1 

\bibitem[\protect\citeauthoryear{Wardzi{\'n}ski et 
al.}{2002}]{2002MNRAS.337..829W} 
Wardzi{\'n}ski, G., Zdziarski, A.~A., Gierli{\'n}ski, et al.\ 2002, MNRAS, 337, 829

\bibitem[Yuan(2001)]{2001MNRAS.324..119Y} Yuan, F.\ 2001, \mnras, 324, 119 

\bibitem[Yuan(2003)]{2003ApJ...594L..99Y} Yuan, F.\ 2003, \apjl, 594, L99 

\bibitem[\protect\citeauthoryear{Yuan et al.}{2007}]{2007ApJ...659..541Y} 
Yuan, F., Zdziarski, A.~A., Xue, Y., \& Wu, X.-B. 2007, ApJ, 659, 541 

\bibitem[\protect\citeauthoryear{Zdziarski \& Gierli\'nski}{2004}]{zg04}
Zdziarski, A. A., \& Gierli\'nski, M. 2004, Progr.\ Theor.\ Phys.\ Suppl., 155, 99

\bibitem[Zdziarski et al.(2002)]{2002ApJ...578..357Z} Zdziarski, A.~A., 
Poutanen, J., Paciesas, W.~S., \& Wen, L.\ 2002, \apj, 578, 357 (Z02)

\bibitem[Zi{\'o}{\l}kowski(2005)]{2005MNRAS.358..851Z} Zi{\'o}{\l}kowski, 
J.\ 2005, \mnras, 358, 851 

\end{thebibliography}
\end{document}